\documentclass[reprint,aps,amsmath,amssymb,twoside,prd,showkeys,superscriptaddress]{revtex4-1}
\pdfoutput=1
\usepackage{verbatim}
\usepackage{comment}
\usepackage{graphicx}
\usepackage[T1]{fontenc}
\usepackage{epsfig}
\usepackage{bm}
\usepackage{amssymb}
\usepackage{float}
\usepackage{amsmath}
\usepackage{subfigure}
\usepackage{dcolumn}
\usepackage{cancel}
\usepackage[colorlinks]{hyperref}
\usepackage[usenames,dvipsnames]{color}
\hypersetup{
     breaklinks=true,
    pdfstartview={FitH},    
    colorlinks=true,       
    linkcolor=blue,          
    citecolor=red,        
    filecolor=magenta,      
    urlcolor=blue,           
    anchorcolor=green,      
    linktocpage=true
}

\def\doi{http://doi.org}

\newcommand{\HCd}{\mathcal{H}}

\def\HCdt0{\tilde{\HCd}_{0}}

\newcommand{\affcam}{DAMTP, Centre for Mathematical Sciences, University of Cambridge, Wilberforce Road, Cambridge CB3 0WA, United Kingdom}
\newcommand{\affkicc}{Institute of Astronomy, University of Cambridge, Madingley Road, Cambridge, CB3 0HA, UK}

\begin{document}

\title{Tracking the Local Group Dynamics by Extended Gravity}

\author{David Benisty}
\email{db888@cam.ac.uk}
\affiliation{\affcam}\affiliation{\affkicc}
\author{Salvatore Capozziello}
\email{capozziello@unina.it}
\affiliation{Dipartimento di Fisica, {\it "E. Pancini"}, Universita' {\it "Federico II"} di Napoli,
Compl. Univ. Monte S. Angelo Ed. G, Via Cinthia, I-80126 Napoli (Italy)}
\affiliation{Scuola Superiore Meridionale, Largo S. Marcellino 10, I-80138 Napoli, Italy}
\affiliation{INFN Sez. di Napoli, Compl. Univ. Monte S. Angelo Ed. G, Via Cinthia, I-80126 Napoli, Italy}

\begin{abstract}
The Local Group (LG) of galaxies, modeled as a two body problem, is sensitive to  cosmological  contributions like those related to the presence of a cosmological constant $\Lambda$ into dynamics. Here we study the LG dynamics in the context of Extended Theories of Gravity like $f(R)$ gravity considered as dark energy and dark matter contributions. In the first approach, we perturb the dark energy effect considering a Yukawa-like interaction that naturally emerges from $f(R)$ gravity in the weak field limit.  {We assume the mass of LG from simulations and, from this, derive constraints on  the Yukawa couplings: $\alpha < 0.581$ and  $m_{grav} < 5.095 \cdot 10^{-26} \, eV/c^2$.} In the second part, considering a minimal extension of General Relativity, i.e. $f(R) \sim R^{1+\epsilon}$, with $|\epsilon|\ll 1$,  we investigate the possibility that it  replaces dark matter as a MOND-like theory. We find that there is a value of the parameter $\beta$ (derived starting from $\epsilon$) which gives a minimal value for the LG mass.  {Moreover, this particular potential allows to calculate the ratio of dark matter and baryonic matter for the LG to be. We show that this ratio could falsify MOND-like theories.}
\end{abstract}
\maketitle
\section{Introduction}
Almost twenty years after the observational evidence of cosmic acceleration the cause of this phenomenon, dubbed 
 "dark energy" (DE), remains an open issue which challenges the foundations of theoretical physics: whenever the DE is the simplest option - a cosmological constant $\Lambda$ or some alternative theory of gravity explaining cosmological dynamics and large scale structure
\cite{Capozziello:2002rd,Nojiri:2010wj,Tsujikawa:2010zza,Capozziello:2011et,Cai:2015emx,Nojiri:2017ncd,Carlesi:2016yas,Odintsov:2020nwm}. Modified theories of gravity are theoretically and observationally appealing, and ever-increasingly sensitive and precise upcoming surveys provide the exciting possibility of robustly testing them against observational data. Among these alternatives,  extensions of General Relativity can be considered a straightforward and natural approach to retain positive results of Einstein theory and eventually to extend it at infrared and ultraviolet scales (see \cite{Capozziello:2011et} for a detailed discussion).

Beside the cosmological systems giving the probes for DE, the LG of galaxies can be used to track DE effects  \cite{Hoffman:2007fu,Erdogdu:2009zb,Eingorn:2012dg,Eingorn:2012jm,Partridge:2013dsa,Gonzalez:2013pqa,McLeod:2016bjk,McLeod:2019cfg}. So far, many tries to test the cosmological constant effect in the LG dynamics have been considered. Here we extend the research for extended gravity models (such as $f(R)$ gravity) where Yukawa-like potentials naturally emerge in the low energy limit \cite{Stabile, Capozziello:2020dvd,Cardone:2011ze}.  {The only consistent $f(R)$ models which evade the solar system constraints have a chameleon mechanism gravity \cite{Brax:2008hh}.} Assuming the LG mass from different simulations, it is possible to  yield bounds for the Yukawa parameters. This research comes into a very large debate related to the dark matter.

In fact, dark matter is one of the most profound unsolved phenomena in modern astrophysics and cosmology. The standard approach describes dark matter as cold massive particles dubbed Weakly Interacting Massive Particles (WIMPs) \cite{Tao:1989xn,Morales:2002ud,Iocco:2012jt,Conrad:2014tla,Rott:2012gh,Baudis:2013eba,DRUKIER:2013lva,daSilva:2014qba,Cui:2015eba,Arcadi:2017kky,Queiroz:2017kxt}, axions \cite{Peccei:1978fx,Davier:1987dv,Murayama:1998jb,Kim:1998sy,Kim:1999ia,vanBibber:2001ud,Geralis:2009ue,Kim:2009xp,Pajer:2013fsa} or very light axion-like particles \cite{Masso:2002ip,Galanti:2019sya,Ertas:2020xcc}. The presence of dark matter in galaxies is observed from different measurements. The basic one is the mismatch between the predicted Keplerian velocity of orbiting stars in galaxies and the measured one \cite{Rubin:1980zd,Begeman:1991iy}. The measured velocity for large distances is approximately constant. In addition, this constant velocity is related to the luminous mass through the Tully-Fisher relation \cite{Zwaan:1995uu,McGaugh:2000sr,TorresFlores:2011uc,McGaugh:2011ac,Chen:2019ftv}.  Other systems where  dark matter is present are the galaxies. Despite of these relevant astrophysical evidences, no new fundamental particle  has been detected till now neither by direct  nor by indirect detection to address the dark matter issue. In this perspective, instead of searching for new matter ingredients, extragalactic and large scale structure dynamics could be addressed by extending gravitational sector (see e.g. \cite{Capozziello:2017rvz,Capozziello:2020dvd, Napolitano:2012fp}).

An important approach in this debate is constituted by the Modified Newtonian Dynamics (MOND) \cite{Milgrom:1983ca,Bekenstein:1984tv,Sanders:2002pf} which is an alternative explanation to the flat rotation curves for galaxies not implying any new matter ingredient. By violating Newton's second law at low accelerations, the Tully-Fisher relation is recovered, without introducing additional dark matter. The Tensor Vector Scalar theories (TeVeS) is a covariant theory, which in the low energy limit, produces MOND \cite{Bekenstein:2004ne}. Specifically, in \cite{Capozziello:2017rvz}, it is  shown that $f(R)$ gravity yields the flat rotation curves of galaxies recovering MOND in the weak field limit. In the second part of this paper, we investigate the LG dynamics, assuming that dark matter is an effect emerging from some $f(R)$ gravity adopting the same philosophy as in \cite{Capozziello:2017rvz} and we show that the LG gives prediction for the ratio between and dark matter and dark energy that could be tested with the observations.

The structure of the paper is the following: Section \ref{sec:form} formulates the timing argument (TA) and the method we use. Section \ref{sec:FRDE} discusses the effect of Yukawa potential on the LG mass and constraints the Yukawa parameters. Section \ref{sec:FRDM} is devoted to modified gravity the capable of mimicking dark matter in the LG system. Here, we calculate the predicted mass of the LG. Section \ref{sec:dis} summarizes the results.

\section{The total Mass from the Timing Argument}
\label{sec:form}
\begin{figure}[t!]
 	\centering
\includegraphics[width=0.49\textwidth]{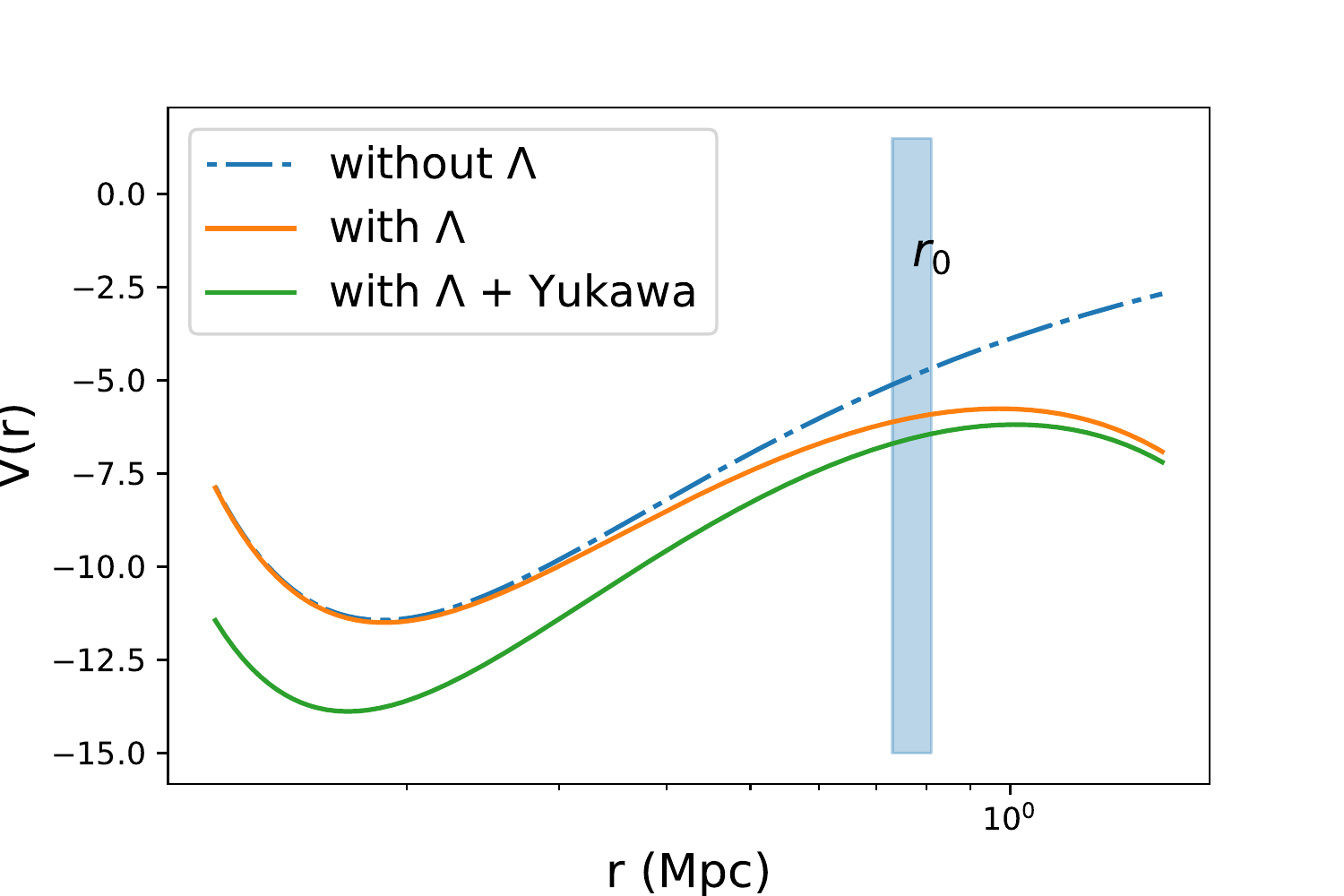}
\caption{\it{ {The effective potential vs. radius for two-body system, assuming $M = 4\cdot 10^{12} \, M_{\odot}$, for different cases: \textbf{dashed - blue:} Newtonian potential, \textbf{orange:} Newtonian potential with Cosmological Constant (from Planck 2018 value), \textbf{green:} Yukawa type potential (with the values $\alpha = 0.1, m_g = 0.01_{Mpc^{-1}}$) and with the Cosmological Constant.}}}
 	\label{fig:potential}
  \end{figure}

The Cosmological Constant  is considered to govern dynamics at cosmological scales \cite{Chernin:2000pq,Baryshev:2000kw,Chernin:2001nu,Karachentsev:2003eh,Chernin:2003qd,Teerikorpi:2005zh,Chernin:2009ms,Chernin:2006dy,Teerikorpi:2010zz,Chernin:2015nna,Silbergleit:2019oyx}. Hence we include the Cosmological Constant contribution in our analysis. The center of mass coordinate system is defined by the relative distance $r$ and the relative velocity $v$. The relative distance variation reads \cite{Emelyanov:2015ina,Carrera:2006im}:
\begin{equation}\label{ENL}
\ddot{r} = \frac{l^2}{r^3}-\frac{GM}{r^2} +  \frac{1}{3}\Lambda c^2 \, r,
\end{equation}
where $\vec{l} =  \vec{r} \times \vec{v}$ and $M$ is the total mass of MW and $M_{31}$. Fig. \ref{fig:potential} presents the corresponding effective potential with and without $\Lambda$. It is possible to see that $\Lambda$ is important even in these short scales \cite{Eingorn:2012dg,Eingorn:2012jm,Partridge:2013dsa,Gonzalez:2013pqa,McLeod:2016bjk,McLeod:2019cfg,Benisty:2019fzt,Lemos:2020vhj}. There are two extreme points: The closer one is a minimum and the other is a maximum. For large distances, the repulsive force arising from the Cosmological Constant makes impossible any bound structure so the Cosmological Constant gives us a constraint on the maximum size of the bound systems.

\begin{figure}
 	\centering
\includegraphics[width=0.47\textwidth]{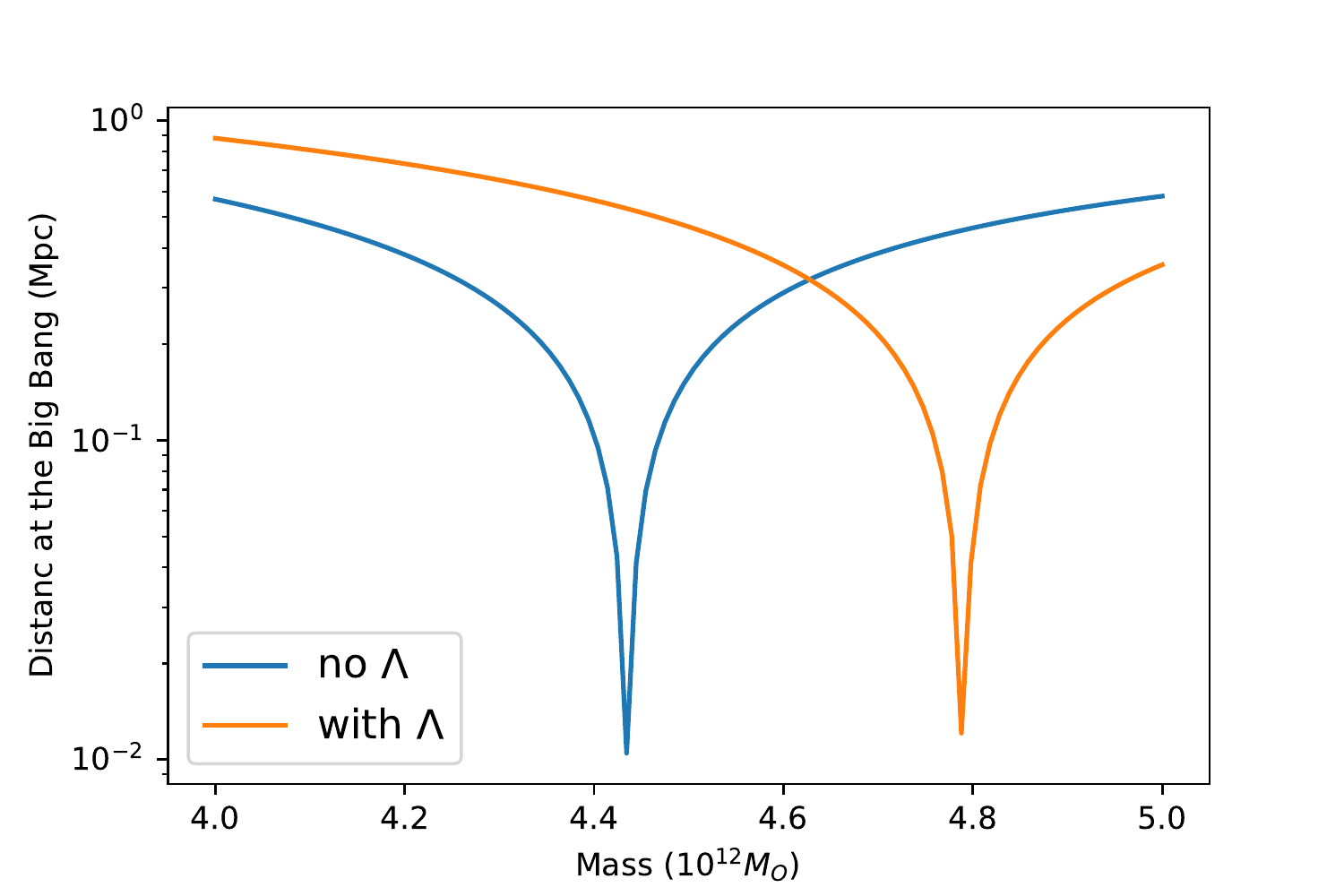}
\caption{\it{ {The evaluated distance between Milky Way and $M\text{31}$ at $t=0$ for different TA masses of the LG. The model predicts the correct mass when $r(t=0) = 0$. The first minimum ($4-5\, 10^{12} M_{\odot}$) points the predicted mass according to the TA. The second minimum points the predicted mass with $\Lambda$. }}}
 	\label{fig:massdistance}
  \end{figure}

The Timing Argument (TA) \cite{1989ApJ345108P,1981Obs101111L} for the Local Group is one of the historical probes for the missing mass problem. The TA assumes that Milky Way ($MW$) and Andromeda ($M31$) have been approaching each other despite cosmic expansion. In its simplest version, the LG roughly consists of the $MW$ and $M31$ as two isolated point masses. Briefly after their formation (for the sake of simplicity the "big bang"),  these two galaxies must have been in the same place with zero separation. Due to the Hubble expansion, these two galaxies moved apart. After a couple of billion years, they slowed down and then moved towards each other again, as a consequence of the gravitational pull. We restrict ourselves to a simple basis of the TA as an isolated, two dimensional system. Galaxies are modeled as point masses. Moreover, as the galaxy pairs are isolated, there are no external gravitational fields. The "initial condition" of the model is $r = 0$, which refers to the "big bang" condition. In order to calculate the mass of the LG, we change the direction of time by considering the opposite direction of the measured velocity now. We evaluate the measured distance of $M31$ to obtain what should be the distance at the "big bang" for different LG masses.

Fig.~\ref{fig:massdistance} presents the distance for the big bang for different masses of LG. When the curve approaches minimum ($r \rightarrow 0$) yielding the predicted mass. The minimum recognizes the mass of the LG as the model predicts. We use the units which can get the current values of the distance and the velocities of $\text{M31}$:
\begin{equation}
\begin{split}
r_{m_{31}} = 0.77 \pm 0.04 \, {\text{Mpc}}, \\ v_{rad} = -109.3 \pm 4.4 \, {km/sec},  \\
v_{tan}(t_u) = 82^{+38}_{-35} \, {km/sec},
\end{split}
\end{equation}
  {based on \cite{vanderMarel:2012xp,2021MNRAS.507.2592S}. Here $v_{rad}$ is the radial velocity towards the $MW$. The cosmological constant is determined by the latest Planck measurements \cite{Aghanim:2018eyx}. When we estimate the initial distance at the big bang, we use the age of the Universe. We use the initial conditions as a Gaussian prior. }

\begin{figure}[t!]
 	\centering
\includegraphics[width=0.39\textwidth]{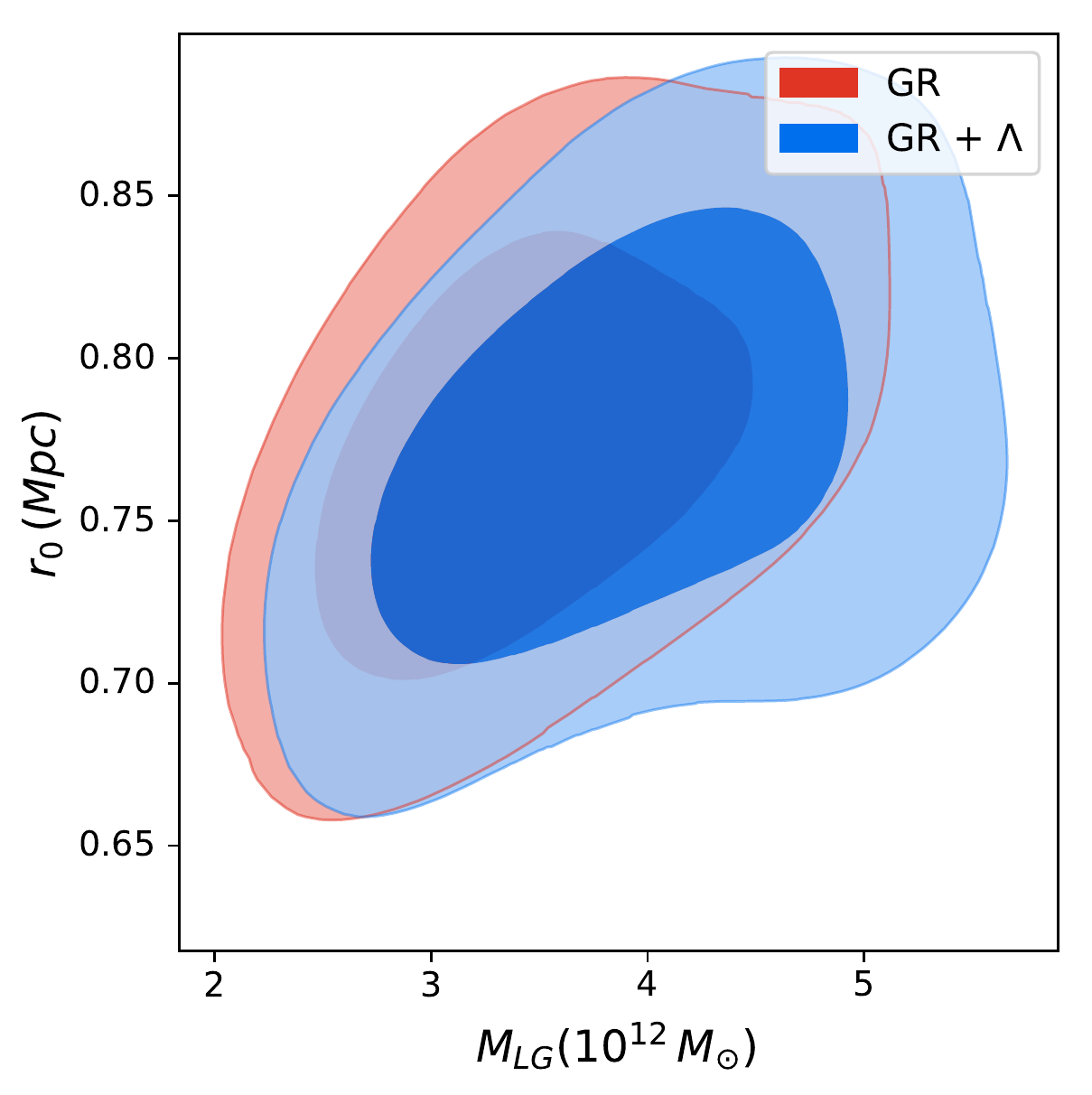}
\caption{\it{The predicted masses from the Timing Argument + Cosmic Bias considering the absence and the  presence the Cosmological Constant. The total mass distribution has a correlation to the distance of $M31$ towards us. Therefore for Yukawa potential that depends  }}
 	\label{fig:massDis}
  \end{figure}

 {Given the simplicity of the Timing Argument and the assumption that the MW and M31 are point masses with constant masses, we have to include some corrections to get unbiased estimates of the true LG mass. This correction arises from other LG members and from the halos around the MW and M31 galaxies \cite{Gupta:2012rh,Bregman:2018dne,Bogdan:2022pbc}. We correct the inferred LG masses with the cosmic bias effect based on \cite{Hartl:2021aio,Benisty:2022ive} with a Gaussian prior of $0.62 \pm 0.2$. The mass of the LG without the Cosmological Constant is: $M_{LG} = \left( 3.28 \pm 0.31\right) \cdot 10^{12} \, M_{\odot},$ and with the Cosmological Constant presence we get: $M_{LG} = \left(3.86 \pm 0.45\right) \cdot 10^{12}  \, M_{\odot}.$ The posterior distributions of the LG mass in presented in Fig~\ref{fig:massDis}. It is possible to see the correlation between the LG mass and the distance of M31 towards us. Therefore for Yukawa type potential that depends on $r$ the predicted mass would be changed and could be constraint based on the history of the LG. }

\section{Yukawa Potential Constraints}
\label{sec:FRDE}
One of the simplest way to test gravity is to constrain the existence of a Yukawa additional term, such that the one particle Newtonian potential assumes the form:
\begin{equation}
\Phi(r) = \frac{G M}{r}\frac{1 + \alpha \, \exp\left[ - m_{grav} r\right]}{1+\alpha},
\end{equation}
where $M$ is the total mass of the system, $G $ is the Newtonian constant and $r$ is the separation between the two objects. For the Yukawa couplings: $\alpha$ is the Yukawa strength and $m$ is the Yukawa mass. For $\alpha$ goes to zero the Yukawa interaction is reduced into the Newtonian one. For $m_{grav}$ goes to zero the Newtonian constant is modified to $G \left(1 + \alpha\right)$. Such a correction arises whenever the force is mediated by a scalar particle of mass $m$. Therefore testing gravity is equivalent to testing the existence of a fifth force of scalar nature. 

Fig~\ref{fig:potential} shows the potential for different values of $\alpha$ and $m_{grav}$. The contribution of the Cosmological Constant effects further the scale of the LG (more then $\sim 1 Mpc$). However, the perturbations of $f(R)$ affect  scales closer than the LG scale. This model cannot replace the Cosmological Constant effect and the latter needs to be added as an additional perturbation.

\begin{figure}[t!]
 	\centering
\includegraphics[width=0.49\textwidth]{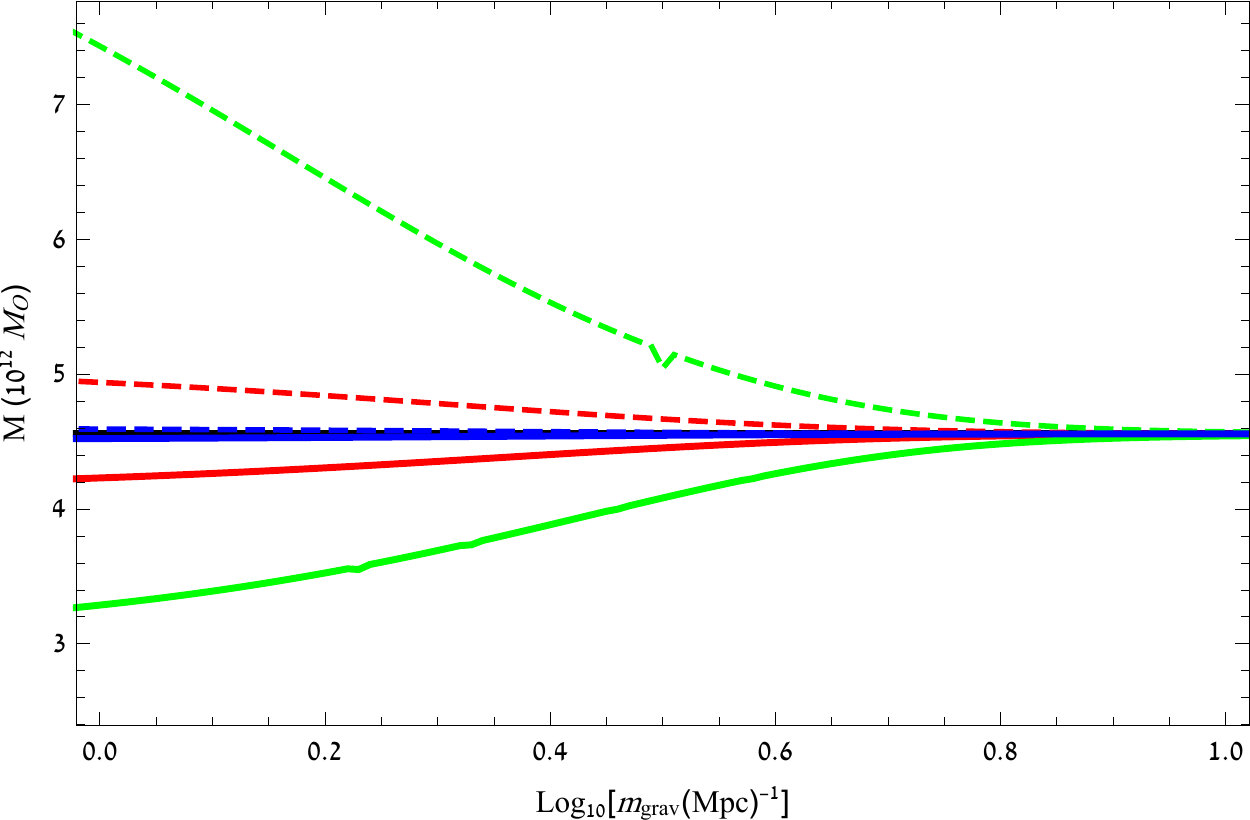}
\caption{{\it{The dependence of the LG mass vs. different values of $\alpha$ and the $m_{\text{grav}}$ in $Mpc^{-1}$. Negative values of $\alpha$ increasing the total mass while positive values of $\alpha$ reducing the total mass. The colors represent different values of $\alpha$: black - $\alpha = 0$, blue - $\alpha = \pm 0.01$, red - $\alpha = \pm 0.1$ and green - $\alpha = \pm 0.5$.}}}
\label{fig:alphaMass}
	\end{figure}

As shown in several studies, Yukawa-like corrections emerges as common  features for several metric theories of gravity \cite{Stabile, Mairi}. In particular for theories containing higher-order curvature invariants or scalar-tensor theories \cite{Capozziello:2010zz}. The field equations for $f(R)$ gravity reads:
\begin{equation}\label{fe}
f'(R)R_{\mu\nu}-\frac{1}{2}f(R)g_{\mu\nu}-f'(R)_{;\mu\nu}+g_{\mu\nu}\Box
f'(R)= T_{\mu\nu}\,,
\end{equation}
with the  trace  
\begin{equation}\label{fetr}
3\Box f'(R)+f'(R)R-2f(R)= T \,.
\end{equation}

where $\kappa$ is the gravitational coupling and $T_{\mu\nu}$ is the matter energy-momentum tensor. We are interested in external solutions so we can ignore the matter contribution.  In the weak field limit, we can perturb the metric tensor with respect to the Minkowski background:
\begin{equation}
g_{\mu\nu}\,=\,\eta_{\mu\nu}+h_{\mu\nu},   
\end{equation}
where $|\eta_{\mu\nu}|\ll |h_{\mu\nu}|$. For an analytic 
 $f(R)$ Lagrangian expandable in Taylor series 
\begin{eqnarray}\label{sertay}
\begin{aligned}
f(R)=\sum_{n}\frac{f^n(R_0)}{n!}(R-R_0)^n\simeq\\
f_0+f'_0R+f''_0R^2+f'''_0R^3+...    
\end{aligned}
\end{eqnarray}
where the prime indicates derivatives with respect to $R$,
the field equations, in the post-Newtonian limit, up to ${\mathcal O}(4)$ order. By adopting the spherical symmetry, the metric can be recast \cite{Capozziello:2012ie,DeMartino:2018yqf,DeLaurentis:2018ahr,DeLaurentis:2018udw,Cardone:2011ze,Napolitano:2012fp,Capozziello:2008ny,Hees:2017aal,Zakharov:2018cbj,Capozziello:2020dvd}:
\begin{equation}
\label{Edd}
    g_{tt}=1-\frac{\Phi(r)}{c^2}\,,\;\;\;\;\;\;  g_{rr}=1+\frac{\Psi(r)}{c^2}\,.
\end{equation}
 {where the functions $g_{tt}$ and $g_{rr}$ give two gravitational potentials:}
\begin{eqnarray}
\label{eq:PHI}\Phi(r) &=& -\frac{G M}{r} \frac{1+\alpha \,   e^{-m_{grav} r}}{1 + \alpha}, \\
\label{eq:PSI}
\Psi(r) &=& -\frac{G M }{r}\frac{1-\alpha \left(1 + m_{grav} r \right) e^{-m_{grav} r}}{1 + \alpha }\,.
\end{eqnarray}
 {The Yukawa strength and scale length read:}
\begin{equation}
\alpha = f_{0}' - 1, \quad m_{grav}^2\,=\,-f'_0/6f''_0,
\end{equation}
 {with a direct relation to the $f\left(R\right)$ function. The potentials coincide for $\alpha \rightarrow 0$ when Newtonian potential is recovered. }

	
In order to test the effect on the LG mass, we use Eq. ~(\ref{ENL}), where the Yukawa potential is replaces the Newtonian potential. Fig~\ref{fig:alphaMass} shows the numerical analysis and the predicted mass for different values of the LG. Negative values of $\alpha$ yields larger mass and positive values of $\alpha$ gives lower values of the mass.

\begin{figure}[t!]
 	\centering
\includegraphics[width=0.48\textwidth]{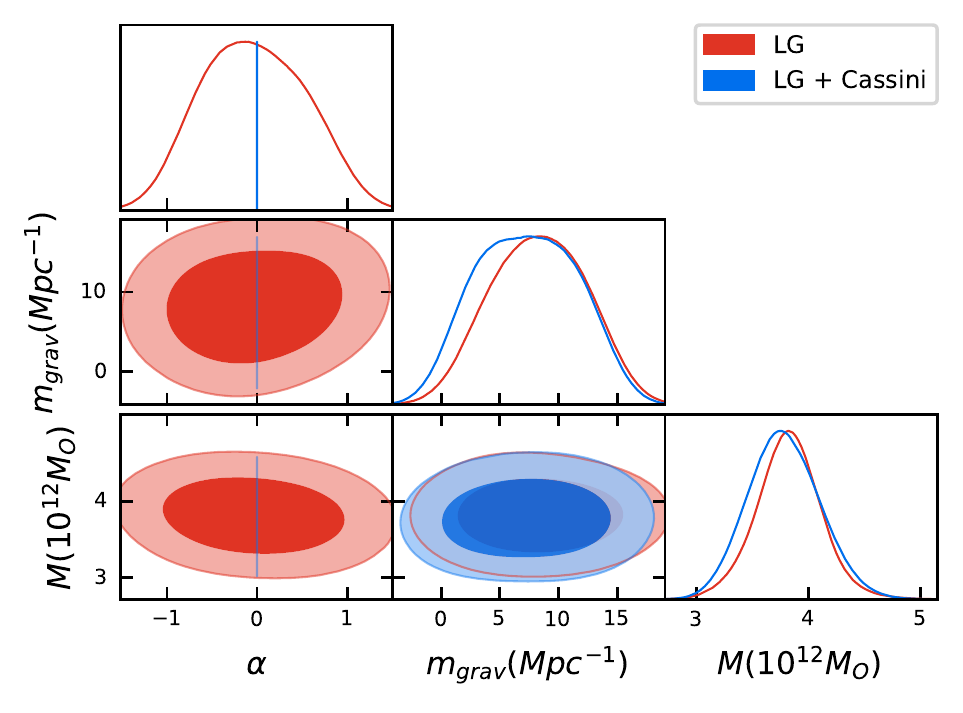}
\caption{\it{The posterior distirubtion for the Yukawa parameters from the Local Group dynamics.   }	}
\label{fig:postFinal}
\end{figure}

In order to constraint the additional parameters we use the predicted mass from the latest LG mass prediction \cite{Benisty:2022ive}. We define the $\chi^2$ difference between the predicted mass and the simulated mass:
\begin{equation}
\chi^2 = \left(\frac{M_{GR + \Lambda} - M(\theta)}{\Delta M_{GR + \Lambda}} \right)^2,
\end{equation}
where $M(\theta)$ is the predicted mass from the set of parameters $\theta = \{r_0, v_r^{0}, v_t^{0}, \Lambda, \alpha, m_{grav} \}$ and $M_{GR + \Lambda}$ is the deduced mass from \cite{Benisty:2022ive}. Regarding the problem of $\chi^2$ minimization, we use nested Markov Chain Monte Carlo sampler as it is implemented within the open-source packaged $\text{Polychord}$ \cite{Handley:2015fda,Lewis:2019xzd}. Fig~\ref{fig:postFinal} shows the posterior distribution for the Yukawa potential. For a model with $m_{grav} = 0$ the largest deviation on $\alpha$ is: 
\begin{equation}
\alpha = \left(2.807 +\- 3.01\right) \cdot 10^{-1},   
\end{equation}
 {or: $\alpha < 0.58$. Assuming the Cassini satellite bound on $\alpha < 2 \cdot 10^{-5}$ with the LG likelihood could gives the constraint on the Yukawa mass:}
\begin{equation}
\begin{split}
 m_{grav} \preceq 5.094 \pm 2.855 \, Kpc^{-1} \\ = \left(3.265 \pm  1.830 \right) 10^{-26} \, eV/c^2.   
\end{split}
\end{equation}
 {The error bars for the parameters are quite large but still indicate a difference from General Relativity. There is some degeneracy between the $\alpha$ and the $\lambda$ parameters that could resolve by assuming a prior on one parameter to get the other. The Yukawa mass is widely discussed in the literature \cite{deRham:2016nuf}. Since the LG scale is about $1 Mpc$. Therefore, it is clear why even the LG system that does not give a strong bound on the $\alpha$ parameter, still gives lower Yukawa mass, that corresponds to the LG large scale. }

 {The study of LG mass could be statistically stronger with another combined systems or with a complete simulation. In order to test the preference of the models we use the Bayesian Evidence (BE). The difference between the BE yields $ \Delta B_{ij} = 0.073 $ which indicates an \textit{indistinguishable preference} for General Relativity.}

\section{Modified Gravity or dark matter}
\label{sec:FRDM}
Let us discuss now whenever $f(R)$ gravity can replace the dark matter in galaxy clusters and in  LG in particular. Ref. \cite{Capozziello:2017rvz} addresses the dark matter in galaxies through the $f(R)$ model of a power\,-\,law case:
\begin{equation}
f(R)  = f_0 R^n\label{eq: frn}
\end{equation}
with $f_0$ a dimensional constant and $n$ any real number.
The choice of  power-law action \eqref{eq: frn} could appear non-natural in order to discuss deviations with respect to General Relativity.  If we assume small deviation with respect to General Relativity, and assume $n=1+\epsilon$ with $|\epsilon| \ll 1$, it is possible to re-write a first-order Taylor expansion as:
 \begin{eqnarray}\label{LOG}
R^{1+\epsilon}&\simeq & R+\epsilon R \,  {\rm log}R +O (\epsilon^2)\,,
 \end{eqnarray}
 and consider small deviations with respect to General Relativity.
The field equations can be set in the Einstein-like form \cite{Capozziello:2011et} as:
\begin{eqnarray}
R_{\mu \nu} - \frac{1}{2} g_{\mu \nu} R& = & \displaystyle{\frac{1}{f'(R)}}
\displaystyle{\Bigg \{ \frac{1}{2} g_{\mu \nu} \left [ f(R) - R
f'(R) \right ] + f'(R)_{; \mu \nu}} \nonumber \\ ~ & - &
\displaystyle{g_{\mu \nu} \Box{f'(R)} \Bigg \}} +
\displaystyle{\frac{T^{(m)}_{\mu \nu}}{f'(R)}} \label{eq:f-var2}\,.
\end{eqnarray}
 Primes denote derivative with respect to $R$. The  terms ${f'(R)}_{; \mu \nu}$ and $\Box{f'(R)}$ give rise to fourth order derivatives in $g_{\mu \nu}$. Taking into account the gravitational field generated by a point like source and solving the field equations (\ref{eq:f-var2}) in the vacuum case, we can write the metric as\,:
\begin{equation}
ds^2 = A(r) dt^2 - B(r) dr^2 - r^2 d\Omega^2 \label{eq: schwartz}
\end{equation}
where $d\Omega^2 = d\theta^2 + \sin^2{\theta} d\varphi^2$ is the spherical line element. Combining the $00$\,-\,vacuum component and the trace of the field equations (\ref{eq:f-var2}) in absence of matter, we get the equation\,:
\begin{equation}
\label{master-low} f'(R) \left ( 3 \frac{R_{00}}{g_{00}} - R
\right ) + \frac{1}{2} f(R) - 3 \frac{f'(R)_{; 0 0}}{g_{00}} = 0 \
.
\end{equation}
Using 
Eq.(\ref{eq: frn}), Eq.(\ref{master-low}) reduces to\,:

\begin{equation}
\label{master-pla} R_{00}(r) = \frac{2 n - 1}{6 n} \ A(r) R(r) -
\frac{n - 1}{2 B(r)} \frac{dA(r)}{dr} \frac{d\ln{R(r)}}{dr} \ ,
\end{equation}
and the trace equation reads\,:

\begin{equation}
\Box{R^{n - 1}(r)} = \frac{2 - n}{3 n} R^n(r) \ . \label{eq:
tracebis}
\end{equation}
For $\epsilon = 0$, Eq.(\ref{eq: tracebis}) reduces to $R = 0$, which, inserted into Eq.(\ref{master-pla}), gives $R_{00} = 0$ and then the  standard Schwarzschild solution is recovered. Expressing $R_{00}$ and $R$ in terms of the metric (\ref{eq: schwartz}), Eqs.(\ref{master-pla}) and (\ref{eq: tracebis}) become a system of differential equations for  $A(r)$ and $B(r)$. A physically motivated 
hypothesis is assuming:
\begin{equation}
A(r) = \frac{1}{B(r)} = 1 + \frac{2 \Phi(r)}{c^2} \label{eq:
avsphi}
\end{equation}
with $\Phi(r)$ the gravitational potential generated by a
point-like mass $m$ at the distance $r$ \cite{Capozziello:2006dp,Zakharov:2006uq}:
\begin{equation}
\Phi(r) = - \frac{G m}{2 r} \left [ 1 + \left ( \frac{r}{r_c}
\right )^{\beta} \right ] \label{equ02}
\end{equation}
\noindent where $r_c$ is an arbitrary parameter, depending on the
typical scale of the considered system and $\beta$ is a universal
parameter:
\begingroup
\setlength{\abovedisplayskip}{0pt}
\setlength{\belowdisplayskip}{0pt}
\begin{equation}
\beta = \dfrac{12 n^2 - 7n - 1 - \sqrt{36 n^4 + 12 n^3 - 83 n^2 + 50n
+ 1} }{6 n^2 - 4n + 2}.
\label{equ03}
\end{equation}
\endgroup
Clearly the gravitational potential deviates from the 
Newtonian one because of the presence of the second term on the right hand side. The critical radius reads:
\begin{equation}
r_c = \sqrt{\frac{G M}{a_0}},
\end{equation}
where $a_0$ is the critical acceleration from MOND $a_0 = 10^{-10} m/s$. The parameter $r_c$ is an integration constant depending on $\beta$. Ref.~\cite{Capozziello:2007wc} shows that it is related to the fact that any $f(R)$ power-law  model is selected by  the existence of Noether symmetries with related conserved quantities. This fact can be considered a physical criterion to retain such models. 

\begin{figure}[t!]
 	\centering
\includegraphics[width=0.51\textwidth]{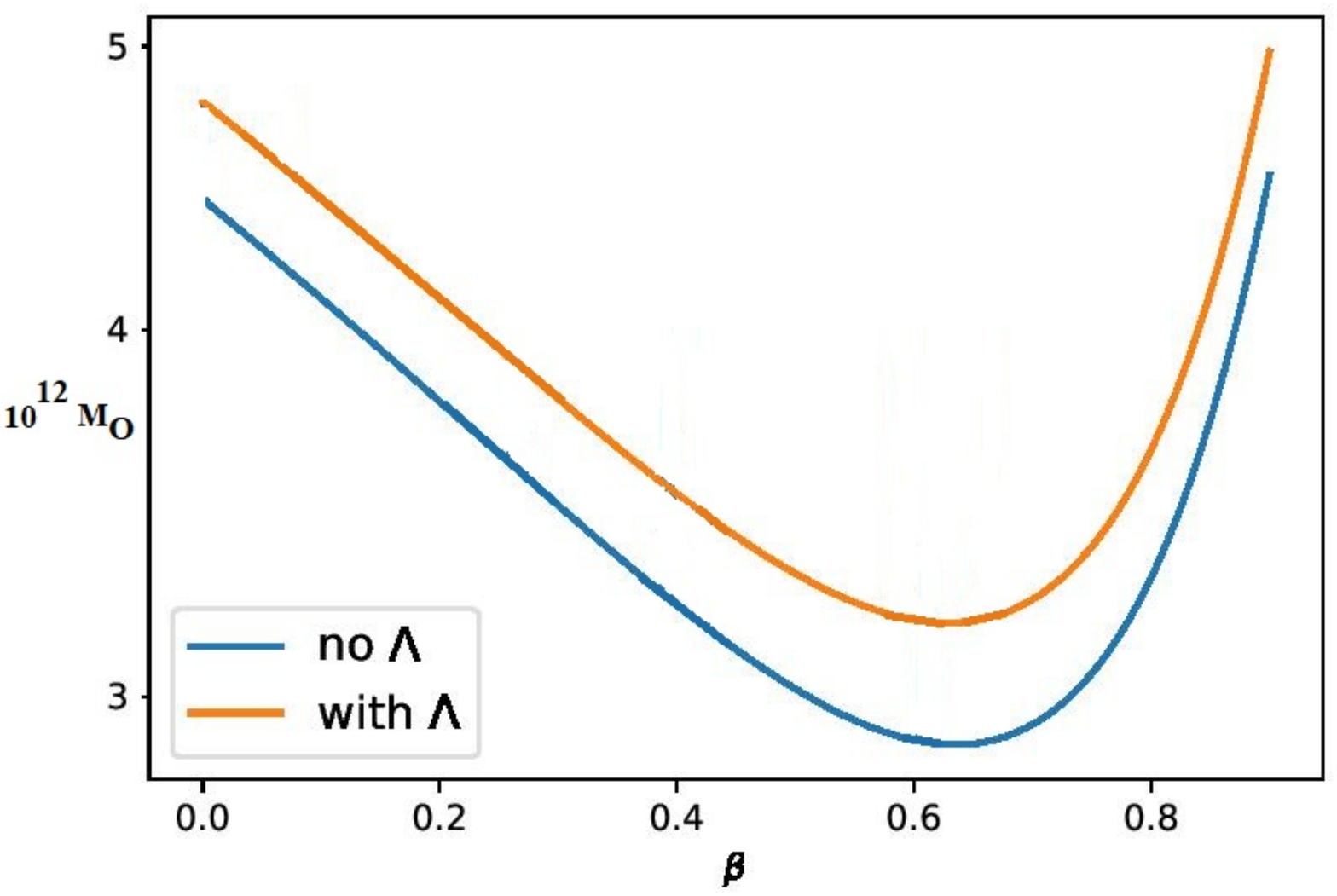}
\caption{\it{The predicted Timing Argument mass of the LG for different values of $\beta$ with and without the Cosmological Constant. When $\beta \sim 0 $ we recover the Newtonian case. For some value of $\beta$ a minimal possible mass is obtained. }}
\label{fig:SpectFullFRDM}
	\end{figure}
\begin{figure}[t!]
 	\centering
\includegraphics[width=0.47\textwidth]{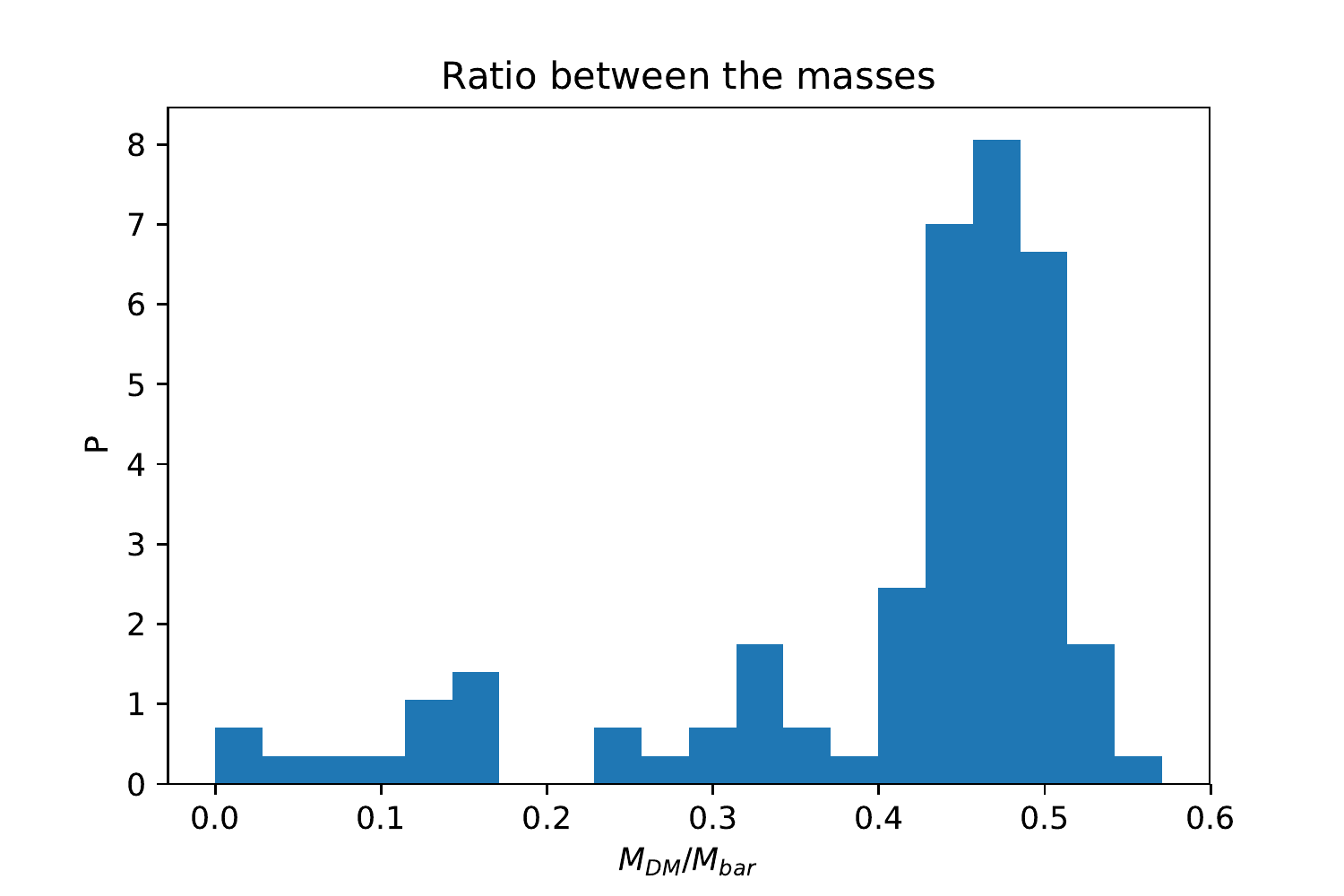}
\caption{\it{The statistical ratio between the the dark matter and the baryonic matter for the LG from the MONDian like potential prediction.}}
\label{fig:motionb}
	\end{figure}
 {It is possible to demonstrate that the potential (\ref{equ02}) can fit  the flat rotation curves in galaxies  like in MOND case \cite{Capozziello:2006dp}. As shown in \cite{Tula1,Tula2}, MOND dynamics can be fully recovered from $f(R)$ gravity. However here we adopt the TA for the potential (\ref{equ02}) for a binary system of galaxies starting from  Eq. (\ref{ENL}). Figure \ref{fig:SpectFullFRDM} shows the predicted mass of LG vs. the parameter $\beta$ that runs from zero to one, with or without a $\Lambda$ term and without the Cosmic Bias factor. The curve shows that there is a minimum predicted mass in a certain value of: }
\begin{equation}
\beta = 0.581^{+0.053}_{-0.081}. 
\end{equation}
For any other value there is a duality for the mass with two different $\beta$ values.  {In order to test the distribution of the predicted mass, we take uniform prior on the parameter $\beta \in [0.;1.]$. The mass of  LG is $\left(2.35 \pm 0.21\right)\cdot 10^{12} \, M_{\odot}$ without $\Lambda$, and $ \left( 2.5 \pm 0.20  \right) \cdot 10^{12} \, M_{\odot}$ with $\Lambda$. }

The MONDian results that we consider here modify the potential while, in the original MOND  model, the formulation modifies the acceleration dependence. Refs.~\cite{McLeod:2019cfg,Benisty:2019fzt,Benisty:2020kys} test different formulations of MOND as a modified acceleration term and find that the prediction for  LG is around ten times lower. While for these formulations, MOND requires additional Past Encounters (PE) to get a reasonable number for the baryonic matter, the $f(R)$ version gives a case closer to the Newtonian one.

The ratio between the masses requires  a different analysis for MOND. In fact, MOND is a formulation that replaces dark matter. So the MONDian TA gives a prediction that all the mass is the baryonic matter alone. The Newtonian case predicts that the baryonic and an additional amount of dark matter constitute the total mass. The dark matter mass is calculated by the difference between the Newtonian prediction and the MONDian prediction. Figure
\ref{fig:motionb} shows the distribution for the ratio between the dark matter and the baryonic matter. The distribution yields the ratio:
\begin{equation}
\frac{Dark Matter}{Baryons} = 0.41^{+0.15}_{-0.38}       
\end{equation}
with $1\,\sigma$ error.  {The ratio between dark matter and baryonic matter in our universe is around six and for the LG is about ten. The fact we get different number for the ratio shows that this function applies for one galaxy, but for two body system there is should be another function. This shows that the TA could test MONDian models of dark matter.}

\section{Discussion and Conclusions}
\label{sec:dis} 
In this paper, we tested the LG of galaxies modeled as a two body problem  sensitive to the cosmological constant ($\Lambda$) contribution considered as an alternative   gravity effect. In the first part, we perturbed the dark energy effect assuming a Yukawa-like correction in the Newtonian potential. Specifically, we derived  the LG mass  from simulations and we constrained  the Yukawa strength. Still we get some ambiguity in the posterior distribution of Yukawa strength vs. the Yukawa length, but we showed that the mass is directly affected from the modified potential.

 {Constraints on $f(R)$ gravity are widely constraint in the literature \cite{Brax:2008hh}, from $\log_{10} f'(R_0) < -4.79$ in galaxy clusters \cite{Cataneo:2014kaa}, $\log_{10} f'(R_0) < -3$ from  GW 170817 \cite{Jana:2018djs} and $f'(R_0) < 3.7 \cdot 10^{-6}$ from the CMB \cite{Boubekeur:2014uaa}. Our constraint is not strong as these constraints ($f'(R_0) < 5.81 \cdot 10^{--1}$), however gives the constraint on $f(R)$ gravity for the first time from the LG dynamics.}

In the second part, we investigate the possibility that a minimal correction to General Relativity, i.e. $f(R) \sim R^{1+\epsilon}$ can replace the effect of dark matter as a MOND-like model. We find that there is a value of the parameter $\beta $, derived by $\epsilon$,  having a minimal value for the mass of  LG. Moreover, this particular potential predicts that the baryonic matter and the dark matter have, more or less, the same order of magnitude. A claim that could be tested with full a N-body simulation. As a conclusion,  we find that the LG dynamics can be considered a good test system for alternative theories of gravity which could be further improved with more realistic simulations and reliable data sets.

In general, modified gravity models of DM usually have some problems in fitting the CMB and other data sets and this is applies to $f(R)$ which mostly can fit galaxy rotation curves. Also in our case it is possible to see that one model of $f(R)$ gives a ratio of $0.41$ for the dark matter and baryonic matter ratio, and in order to fit for the experimental ratio the function possibly needs to be generelized.

\acknowledgments
D.B. thanks the Rothschild and the Blavatnik Fellowships for generous supports. This work has been supported by the European COST actions CA18108 and CA21136 and the research grants KP-06-N58/5. S.C. acknowledges the support of {\it Istituto Nazionale di Fisica Nucleare (INFN) iniziative specifiche QGSKY} and {\it MOONLIGHT2}.

\bibliography{ref.bib}
\bibliographystyle{apsrev4-1}

\end{document}